\documentclass[a4paper,11pt]{article}
\usepackage[utf8]{inputenc}
\usepackage[T1]{fontenc}
\usepackage{authblk}
\usepackage{graphicx}
\usepackage{amsmath}
\usepackage{url}

\begin{document}
\title{On the interplay between mobility and hospitalization capacity during the COVID-19 pandemic: The SEIRHUD model}
\author[1,2,3,4]{Tom\'as Veloz}
\author[1]{Pedro Maldonado}
\author[1]{Samuel Ropert}
\author[1,5]{C\'esar Ravello}
\author[1,5]{Soraya Mora}
\author[1]{Alejandra Barrios}
\author[1]{Tom\'as Villaseca}
\author[1]{C\'esar Valdenegro}
\author[1,5,6]{Tom\'as Perez-Acle}
\affil[1]{Computational Biology Lab (DLab), Fundación Ciencia \& Vida, Santiago, Chile}
\affil[2]{Fundación para el Desarrollo Interdisciplinario de la Ciencia, la Tecnología y las Artes - DICTA}
\affil[3]{Centre Leo Apostel for Interdisciplinary Studies, Vrije Universiteit Brussel}
\affil[4]{Universidad Andr\'es Bello, Facultad de Ciencias para la Vida}
\affil[5]{Universidad San Sebasti\'an}
\affil[6]{Centro Interdisciplinario de Neurociencia de Valpara\'iso, Universidad de Valpara\'iso}
\date{June 2020}

\maketitle
\begin{abstract}
Measures to reduce the impact of the COVID-19 pandemic require a mix of  logistic, political and social capacity. Depending on the country, different approaches to increase hospitalization capacity or to properly apply lock-downs are observed. In order to better understand the impact of these measures we have developed a compartmental model which, on the one hand allows to calibrate the reduction of movement of people within and among different areas, and on the other hand it incorporates a hospitalization dynamics that differentiates the available kinds of treatment that infected people can receive. By bounding the hospitalization capacity, we are able to study in detail the interplay between mobility and hospitalization capacity.
\end{abstract}

\section{Introduction}
The SARS-Cov2 virus, first reported in China by the end of the year 2019, generated a pandemic with a high death toll, having a large impact in the global economy and in the daily lives of millions of people. While a small proportion of countries are experiencing a reduction in the number of new daily cases, the disease is still rapidly spreading in the majority of countries. Due to the lack of innate immunity by humankind and since there is still no cure nor a vaccine, the only way to prevent contagion is to reduce the chances of viral transmission by reducing person to person contact: a strategy generally termed non-pharmaceutical intervention~\cite{Eubank:2020aa}. In this regard, defining strategies to control the spread of the disease while at the same time preserving the well-being of people, preventing the loss of jobs and limiting the impact on the economy, has become an urgent endeavour. Without proper tools to forecast accurately the outcome of possible counter-measures focused on controlling the virus spread, reducing the impact over the sanitary system, the design and execution of optimal strategies is a complex endeavour.

Among all the possible variables that can modulate the impact of the disease, reductions of mobility and increments on the hospitalization capacity are possibly the two most significant ones. On the one hand, it is well known that the case fatality rate (CFR), i.e. the ratio between the COVID-19 deaths and COVID-19 infected people, in the absence of a sanitary system is around $15\%$, and in the presence of a virtually unlimited sanitary system, meaning that every infected person who needs treatment will get it, CFR should be around $2.5\%$~\cite{lauer2020incubation}. As it is also known that the infection peak can reach
 very large values, hence hospitalization capacity can collapse increasing the CFR to values that approximate $15\%$.  On the other hand, reductions of mobility might have different impacts, but also different logistic costs, ranging from total reduction of mobility, i.e. total lock-down, vanishing the disease between two to four weeks, but being impossible to apply because the government would have to manage the biological, psychological and economical necessities of the entire locked-down population, to more realistic reductions such as cancelling massive events, implementing work-from-home, and lock-downs that allow a certain degree of mobility (to buy supplies, workout and walk pets, among others), that move the peak of infection to the future, and also reduce the number of infected people in the peak, flattening the infection curve~\cite{matrajt2020evaluating,alvarez2020simple} \cite{walker2020report}. The increase of the hospitalization capacity and the reduction of mobility are both fundamental aspects of the policies implemented by most countries to control the impact of the COVID-19 pandemic~\cite{das2020critical,verity2020estimates}.   

Since either increasing the hospital capacity or implementing effective reductions of mobility depend on the economic, political and social capabilities of each country, we performed a study to get insights on the interplay between these two variables, so we can identify optimal policies involving these variables but considering the country's capabilities to achieve the proposed goal. To do so, we have extended the classical SEIR model to incorporate four types of infected individuals: asymptomatic, mild symptoms, severe symptoms and critical symptoms. While the first two kinds do not require hospitalization, they are the most infectious due to their ability move without restrictions among the population. In contrast, the other two kinds of infected individuals require hospitalization. Individuals with severe symptoms will require intensive care treatment without the use of mechanical ventilators, while critical individuals will require the use of ventilator \cite{liu2020viral}. To account for these differences, we defined two kinds of hospitalization stages, incorporating intermediate hospitalization states to include facts such as the aggravation of symptoms from severe to critical while hospitalized. Importantly, we assume a finite hospitalization capacity, that can be incremented over time.

Another aspect in which we improve the SEIR model is the incorporation of a mobility parameter to model the reduction of interactions due to changes in social behavior, quarantines, lock-downs, and check-points among others. We start from a mobility baseline which we can reduce by hand to simulate different lock-down scenarios, and also we can set it to the reductions of mobility observed by geo-referenced tracking of anonymized mobile phones data~\cite{google-mobility}. Having both the mobility and hospitalization parameters in hand, we will investigate whether and when the hospitalization system is capable to treat the incoming infected people by calculating the ratio between deaths and infected people who require hospitalization to survive. We termed this parameter as the {\it should-be-hospitalized fatality rate} (SHFR).  

The paper is organized as follows. We first introduce the SEIR model that incorporates mobility. Next, we introduce the extended model with different infection and hospitalization states. Next, we provide numerical examples of the model to understand some interesting qualitative aspects of it, and show how this model forecasts the situation in Chile. We conclude discussing what improvements are still required in the model to achieve more exact predictions of the future of the disease at a country-wide scale.

%Of note, by a collaborative project with the Chilean Ministry of Science %and Technology, our model has been successfully used by the Chilean %government to forecast the dispersion of the COVID19 in both the %population and the sanitary system, and also to evaluate the impact of %the mitigation strategies currently in place in Chile. 

\section{Models}
\subsection{A SEIR model with mobility}

The SEIR epidemiological model, and several variations of it, have been extensively studied both for the purpose of understanding dynamics~\cite{al2012modeling,lyra2020covid,roda2020difficult,liu2020viral,picchiotti2020covid,linka2020outbreak,lopez2020modified} as well as optimal policy making~\cite{chandak2020epidemiologically,askitas2020lockdown,pan2020association,fang2020transmission,hou2020effectiveness}. For a comprehensive review see~\cite{berger2020seir,park2020systematic}. 

Let $N$ be the size of a population where no birth occurs (or recently born babies are not susceptible to COVID19). We define $S,E,I$, and $R$ as the amount of susceptible, exposed, infected and recovered inhabitants respectively, with $S+E+I+R=N$.

We shall now describe the interactions between susceptible and infected populations by assuming two general conditions
\begin{itemize}
    \item There is an intrinsic rate infection parameter $\beta$ which modulates the success of infection transmission due to interactions, and does not depend on the population sizes or time, but only on the virus type. We assumed $\beta=0.2$ obtained from literature~\cite{park2020systematic}.
    
    \item The rate of interaction events is proportional to the product between the susceptible population $S$, the effective density of infected people given by $\frac{I}{N}$, and a factor $\alpha(t)$ which denotes changes in mobility, and thus the number of interactions, due to personal, social, and political measures that change the behaviour of people.
    \end{itemize}
The latter specification is only an approximation of the real population dynamics because we are implicitly assuming random mixing, thus neglecting the spatial constrains that are imposed on the movement of people on a more realistic scenario. Therefore our mobility parameter, on the one hand accounts for the fact that space reduces the number of interactions compared to random mixing when the population density is sufficiently large, and on the other hand it allows to introduce the effect of the mobility reductions produced by quarantines, lockdowns, personal mobility restrictions, etc. Hence, our approximation is useful to not only describe the qualitative dynamics but also to provide estimations of the total number of infected people, the one or many peaks, and various other measures of the pandemics such as SHFR. We elaborate further on the mobility issue in the discussion section.

Using the standard construction of the SEIR model for $E,I,$ and $R$, we have that the equations ruling the evolution of this system are 

\begin{equation} \label{eu_eqn1}
\frac{dS}{dt} = -\alpha(t)\beta {S}\frac{I}{N}
\end{equation}
\begin{equation} \label{eu_eqn}
\frac{dE}{dt} = \alpha(t)\beta {S}\frac{I}{N} - \sigma E 
\end{equation}
\begin{equation} \label{eu_eqn}
\frac{dI}{dt} = \sigma E - \gamma I
\end{equation}
\begin{equation} \label{eu_eqn}
\frac{dR}{dt} = \gamma I
\end{equation}
\begin{equation} \label{eu_eqn}
\end{equation}
Where  $\mathbf{\gamma ^{-1}}$ and $\mathbf{\sigma ^{-1}}$ are the recovery time and incubation period.

Since our main interest is the interplay between mobility and hospitalization capacity, we extended this SEIR model to appropriately incorporate the hospitalization phenomenology.

\subsection{Incorporating different types of infected states and the hospitalization system}

Before explaining in deeper details the extended transitions of our model, it is important to keep in mind that we have four kinds of infected people. While the asymptomatic and mild symptoms, $I_{as}$ and $I_{mi}$ respectively, do not require hospitalization before recovering (and thus only contribute to the transmission dynamics), infected with severe and critical symptoms, $I_{se}$ and $I_{cr}$ respectively, require hospitalization with different treatment. Infected with severe symptoms require intensive care and possibly the use of oxigen, and infected with critical symptoms will require ventilator treatment. Since the treatment of severe patients is limited by the number of hospitalization UCI beds, and the treatment of critical patients is limited by the number of ventilators, we consider a maximal (but time dependent) UCI-hospitalization (hospitalization from now on) and ventilator capacities $H_{tot}$ and $V_{tot}$, respectively. Whenever the number of patients requiring each of this treatments reach its maximal capacity, the new infected people coming cannot receive the respective treatment and eventually die. Considering that the death rate is $2.5\%$ and $15\%$ with and without treatment capacity, it is extremely important understand the conditions under which such capacity is reached.

\begin{figure}[h!]
    \centering
    \includegraphics[height=6.5cm,width=12.8cm]{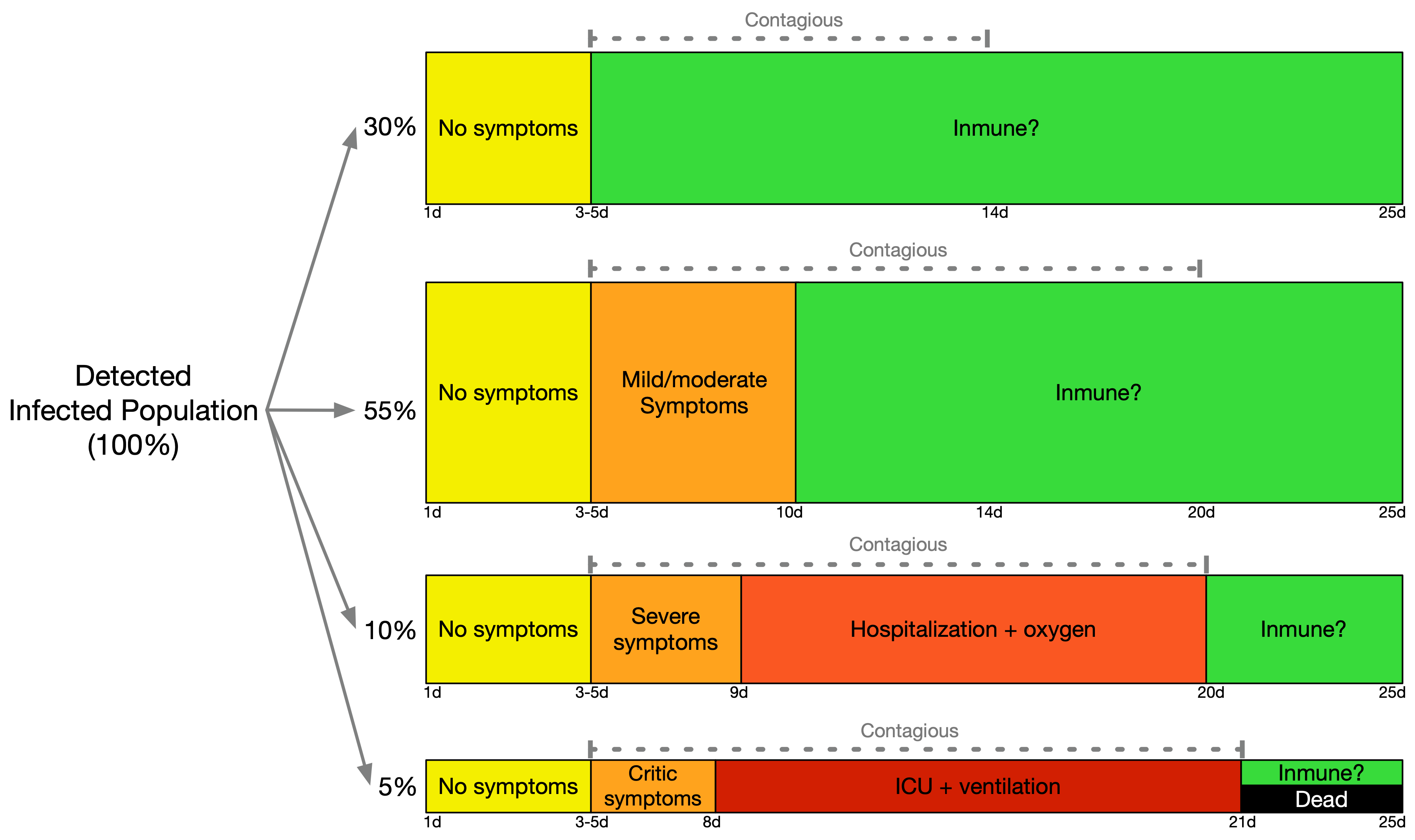}
    \caption{Covid19 states and the distribution of infected population detected, according to the severity of their symptoms. Namely, $ 30 \% $ of infected people will show no symptoms, while $ 55 \% $ of those will have mild symptoms. In both cases, hospitalization will not be required. The $ 15 \% $ of those infected will present severe symptoms, so they will need hospital care. The $ 5 \% $ of those infected will show severe symptoms, requiring hospitalization via ICU / UTI coupled with mechanical respirator. Of these latest patients, around $ 50 \% $ dies from complications derived from Covid19.}
    \label{infection-path}
\end{figure}

\begin{figure}[h!]
    \centering
    \includegraphics[height=7.0cm,width=12.8cm]{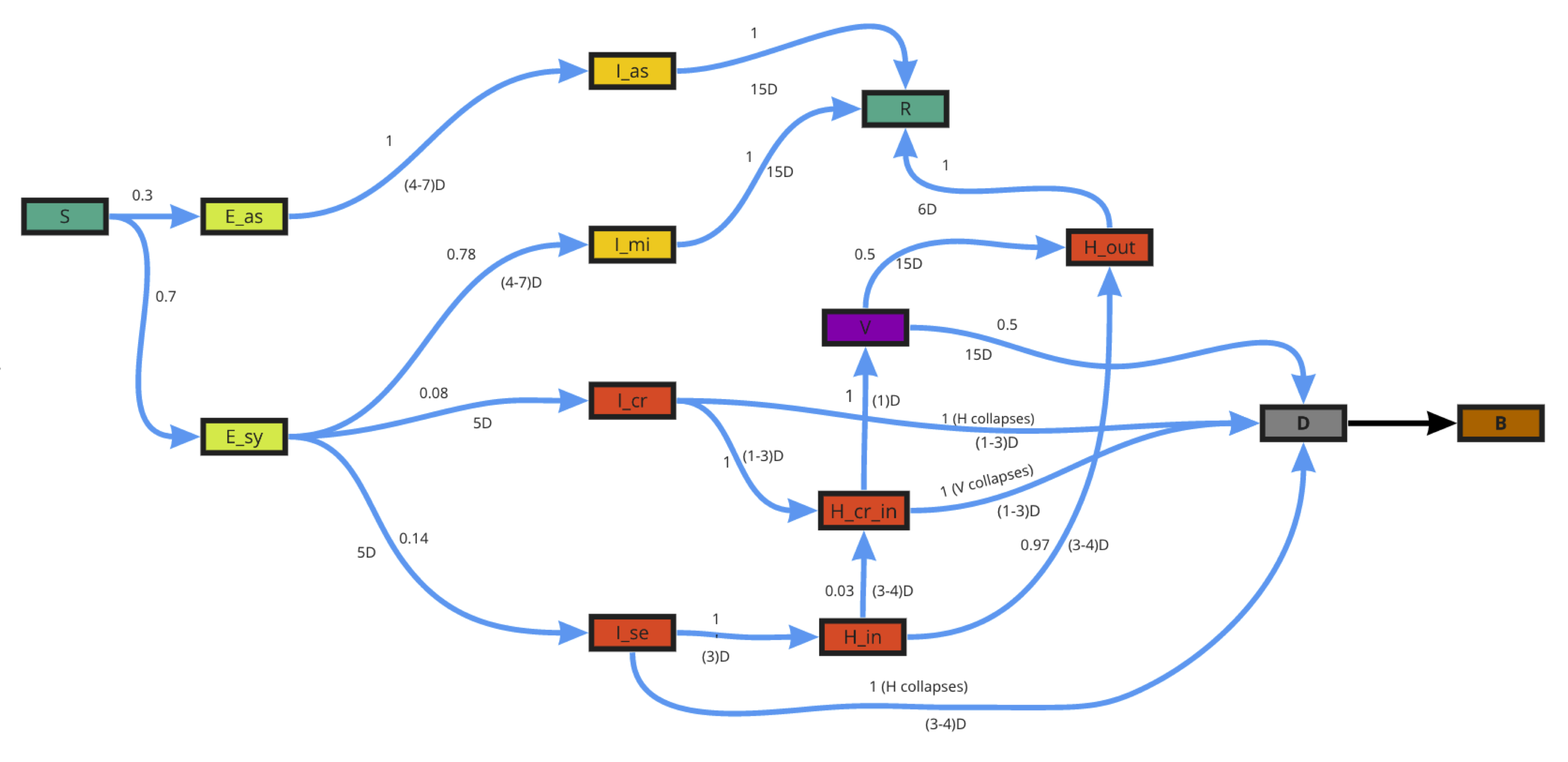}
    \caption{Representation of the transitions in the SEIRHUD model. For a given transition $x\to y$, the number above the transition arrow represents the proportion of population who undergoes such transition from the previous state, and the number below the arrow is an estimate of days in which such transition is expected to occur. Hence the ratio between the first and second number is the transition rate $\sigma_{x,y}$.}
    \label{SEIRHUD}
\end{figure}

In order to facilitate the understanding of the situation we provide two figures. In the fig.~\ref{infection-path} we show a diagram of the time and proportions of the different transitions compiling various sources in the literature~\cite{verity2020estimates,lauer2020incubation,liu2020viral,ferguson2020impact,park2020systematic,giordano2020modelling,weissman2020locally}, and in fig.~\ref{SEIRHUD} we show a compartmental-model-diagram of those transitions, making more explicit the way in which the differential equations are built.

We will explain in more detail the transitions of the model in  fig.~\ref{SEIRHUD}:

\begin{itemize}
    \item Every susceptible $S$ can become exposed with an asymptomatic or symptomatic future. We introduce two exposed states $E_{as}$ and $E_{sy}$ to differentiate them.
    \item Every $E_{as}$ becomes infected asymptomatic $I_{as}$, which later becomes recovered $R$.
    \item Every $E_{sy}$ can become either infected with mild symptoms $I_{mi}$, infected with severe symptoms $I_{se}$ or infected with critical symptoms $I_{cr}$. While $I_{mi}$ will not require hospitalization, $I_{se}$ will require basic hospitalization to recover (at most oxigen), and $I_{cr}$ will require the use of a ventilator to recover.  
    \item Since not only the proportion but also the mobility of $I_{se}$ and $I_{cr}$ is small compared to $I_{as}$ and $I_{mi}$, we consider that the infection rate depends only on $I_{as}$ and $I_{mi}$.
    
    \item When $I_{se}$ is hospitalized, then it enters in a state $H_{in}$ which means it is hospitalized. From such state it can evolve to either $H_{out}$ meaning it will recover, or to $H_{cr-in}$ meaning its symptoms worsen so it requires ventilation treatment. 
    \item If the hospitalization capacity is reached, then $I_{se}$ cannot be hospitalized and dies.
    \item When $I_{cr}$ is hospitalized it enters in state $H_{cr-in}$ and goes to the ventilator treatment state $V$. From $V$, it can either improve and transit to $H_{out}$ ending up as recovered $R$, or die during the treatment.
    \item  If the hospitalization capacity is reached, then $I_{cr}$ and $I_{se}$ cannot be hospitalized (represented by states $H_{cr-in}$ and $H_{in}$ respectively), and eventually transit to a dead state. 
    \item For both $I_{cr}$ and those $I_{se}$ whose symptoms worsen, once hospitalized in state $H_{cr-in}$, if the ventilator capacity is reached, then  they cannot enter in the ventilator treatment and die.
\end{itemize}

The system is ruled by the following equations

\begin{equation}
\begin{split}
    S'&=-\beta \alpha(t) \frac{S}{N}(I_{as}+I_{mi})\\
E_{as}'&= \sigma_{S,{E_{as}}}\beta \alpha(t) \frac{S}{N}(I_{as}+I_{mi})-\sigma_{E_{as},I_{as}}E_{as}\\
E_{sy}'&= \sigma_{S,{E_{as}}}\beta \alpha(t) \frac{S}{N}(I_{as}+I_{mi})-(\sigma_{E_{sy},I_{mi}}+\sigma_{E_{sy},I_{se}}+\sigma_{E_{sy},I_{cr}})E_{sy}\\
I_{as}'&= \sigma_{E_{as},I_{as}}E_{as}-\sigma_{I_{as},R}I_{as}\\
I_{mi}'&= \sigma_{E_{sy},I_{mi}}E_{sy}-\sigma_{I_{mi},R}I_{mi}\\
I_{se}'&=\sigma_{E_{sy},I_{se}}E_{sy}-\sigma_{I_{se},D}H_{sat}-\sigma_{I_{se},H_{in}}(1-H_{sat})\\
I_{cr}'&=\sigma_{E_{sy},I_{cr}}E_{sy}-\sigma_{I_{cr},H_{cr}}(1-H_{sat})-\sigma_{I_{cr},D}H_{sat}\\
H_{in}'&=\sigma_{I_{se},H_{in}}(1-H_{sat})-\sigma_{H_{in},H_{cr}}H_{in}-\sigma_{H_{in},H_{out}}H_{in}\\
H_{cr}'&=\sigma_{I_{cr},H_{cr}}I_{cr}(1-H_{sat})+\sigma_{H_{in},H_{cr}}H_{in}-\sigma_{H_{cr},V}H_{cr}(1-V_{sat})-\sigma_{H_{cr},D}H_{cr}V_{sat}\\
V'&=\sigma_{H_{cr},V}H_{cr}(1-V_{sat})-\sigma_{V,H_{out}}V(1-H_{sat})-\sigma_{V,D}V\\
H_{out}'&=\sigma_{H_{in},H_{out}}H_{in}+\sigma_{V,H_{out}}V(1-H_{sat})-\sigma_{H_{out},R}H_{out}\\
D'&=\sigma_{I_{cr},D}H_{sat}+\sigma_{I_{se},D}H_{sat}+\sigma_{H_{cr},D}H_{cr}V_{sat}+\sigma_{V,D}V\\
R'&=\sigma_{I_{as},R}I_{as}+\sigma_{I_{mi},R}I_{mi}+\sigma_{H_{out},R}H_{out},
\end{split}
\end{equation}
where the transition parameters $\sigma_{x,y}$ represent the rate of the transition\footnote{the ratio between numbers above and below the arrow representing the transition in fig.~\ref{SEIRHUD}.} from state $x$ to state $y$, $H_{sat}$ is a function whose value is $1$ if $H_{in}+H_{cr}+H_{out}\geq H_{tot}$ and $0$ else, and $V_{sat}$ is a function whose value is $1$ if $V\geq V_{tot}$ and $0$ else.

The code for solving and analyze these equations is available for download in~\cite{githuburl}. The parameter table is below

\begin{table}[h!]
\begin{center}
\begin{tabular}{|c|c|c|}\hline
 rate & value & meaning \\ \hline \hline 
 $\beta$ & 0.19 & success of transmission infected to susceptible \\  \hline
$\sigma_{E_{as},I_{as}}$ & $0.2$ & change from exposed to infected asymptomatic\\ \hline   
$\sigma_{E_{sy},I_{mi}}$ & $0.156$ & change from exposed to infected mild symptoms \\ \hline   
$\sigma_{E_{sy},I_{se}}$ & $0.028$ & change from exposed to infected severe symptoms \\ \hline   
$\sigma_{E_{sy},I_{cr}}$ & $0.016$ & change from exposed to infected critical symptoms \\ \hline   
$\sigma_{I_{as},R}$ & $0.066$ & change from infected asymptomatic to recovered \\ \hline   
$\sigma_{I_{mi},R}$ & $00.066$ & change from infected asymptomatic to recovered \\ \hline   
$\sigma_{I_{se},H_{in}}$ & $1$ & change from infected severe symptoms to hospitalized \\ \hline
$\sigma_{I_{cr},H_{cr-in}}$ & $1$ & change from infected critical symptoms to hospitalized \\ \hline
$\sigma_{I_{se},D}$ & $0.33$ & change from infected severe symptoms to death \\ \hline
$\sigma_{I_{cr},D}$ & $0.33$ & change from infected critical symptoms to death \\ \hline
$\sigma_{H_{in},H_{out}}$ & $0.2475$ & change from hospitalization state to start of the recovery \\ \hline
$\sigma_{H_{in},H_{cr-in}}$ & $0.01$ & worsen of symptoms during hospitalization\\ \hline
$\sigma_{H_{cr-in},V}$ & $1$ & change from hospitalization state to ventilation \\ \hline
$\sigma_{H_{cr-in},D}$ & $0.33$ & change from hospitalization to death \\ \hline
$\sigma_{V,D}$ & $0.03$ & change from ventilator to death \\ \hline
$\sigma_{V,H_{out}}$ & $0.03$ & change from ventilator to start recovering \\ \hline
$\sigma_{H_{out},R}$ & $0.33$ & exit from hospitalization \\ \hline
\end{tabular}
\caption{Parameters we use to run our solver.}
\label{parameter-table}
\end{center}
\end{table}

\section{Preliminary Analysis}

We will first study the interplay between the mobility function $\alpha(t)$ and the hospital capacities $H_{tot}$ and $V_{tot}$ using a numerical scenario. %and later we will fit and forecast the situation in the metropolitan region of Chile, which is the region that concentrates the majority of population, infected cases, and hospital capacity. 

We will concentrate in the SHFR defined by the ratio between the number of dead people by COVID19 and the number of infected people that require hospitalization.
\begin{equation}
    \text{SHFR}(t)=\frac{D(t)}{\displaystyle\int_{0}^{t}I_{se}(\tau)+I_{cr}(\tau) d\tau}
\end{equation}
From Fig.~\ref{infection-path} we have that $15\%$ of the infected have either severe or critical symptoms, and in case there is no hospitalization capacity whatsoever they would all die. In this case $\text{SHFR}(t)=1$. Moreover, since we have that the fraction of people with severe symptoms is twice the fraction of people with critical symptoms, we have that $I_{se}\sim 2I_{cr}$, and that the chance of recovering of hospitalized patients with severe symptoms is almost $100\%$, while for the case of critical patients in ventilators, the chance of recovering is $50\%$, we can estimate a lower bound of the $\text{SHFR}(t)$ as follows:
\begin{equation*}
\begin{split}
    \text{SHFR}(t)&=\frac{D(t)}{\displaystyle\int_{0}^{t}I_{se}(\tau)+I_{cr}(\tau) d\tau}\\
    &=\frac{\displaystyle\int_{t=0}^{\tau}\epsilon I_{se}(\tau)+0.5 I_{cr}(\tau) d\tau}{\displaystyle\int_{t=0}^{t}I_{se}(\tau)+I_{cr}(\tau) d\tau}\sim\frac{\displaystyle\int_{t=0}^{t}I_{cr}(\tau)dt}{\displaystyle\int_{t=0}^{t}I_{cr}(\tau)dt}\left(\frac{2\epsilon+0.5}{2+1}\right)\sim \frac{1}{6}.
\end{split}
\end{equation*}
% ¿No es esto darse muchas vueltas para llegar a que el limite inferior de la SHFR debería ser la tasa de letalidad?
Hence, we have that $\text{SHFR}(t)$ in active pandemic conditions, i.e. if none of the measured variables are zero, should range between an optimal value $\frac{1}{6}$ and a worst value $1$. 

In our simulation we consider a system with $1.000.000$ people, and leave the $\alpha$ parameter constant through each simulation at different values from $0.2$ to $1$. Based on~\cite{bed-data-world}, we consider an estimate between 1 and 25 hospitalization beds per thousand people, covering most of the world's realities, and estimate that $30\%$. of such capacity corresponds to ventilators. This estimation can be considered rough, however, it is still useful for the purpose of understanding the interplay between hospitalization capacity and mobility. 

\begin{figure}[h!]
    \centering
    \includegraphics[height=6.5cm,width=8.5cm]{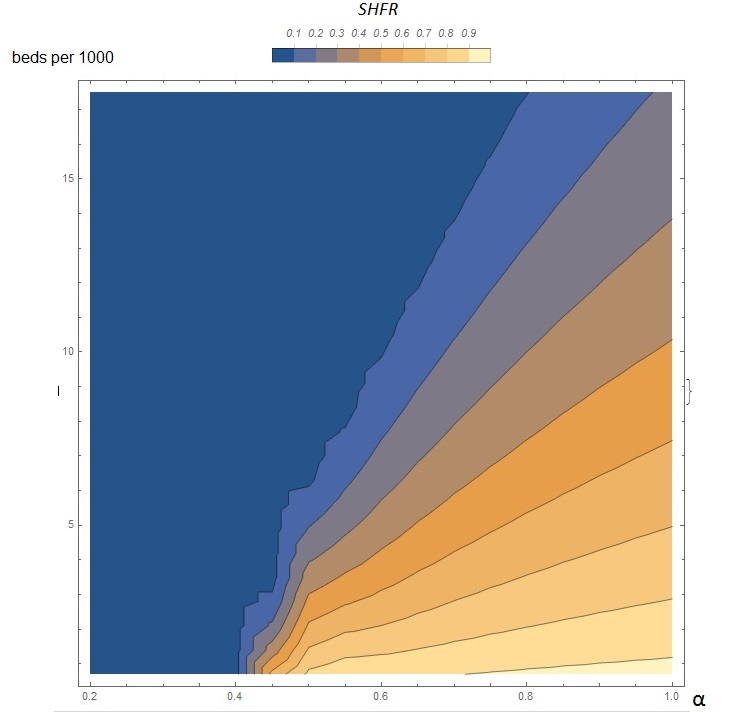}
    \caption{SHFR value for different values of $\alpha$ and $H_{tot}$.}
    \label{SHFRStatic}
\end{figure}

In figure~\ref{SHFRStatic} we observe the converging SHFR value for the tested scenarios (convergence found to be stable after 250 days in all scenarios). It is interesting to note that in this case the mobility $\alpha\leq 0.4$ controls the pandemics in all possible $H_{tot}$ scenarios. This result is consistent with~\cite{xiong2020simulating} where a similar mobility parameter was found. This means that with mobility below $0.4$ there is no spread of the disease, i.e. $R_0<1$. For $\alpha>0.4$ we observe that in the absence of beds very little mobility induces a big increase in SHFR. For example in the bottom case (1 bed per thousand people), changing the mobility from $\alpha=0.4$ to $0.5$ we increase SHFR from $0.2$ to $0.8$, while the same change of mobility having 3 beds per thousand people increases SHFR up to $0.4$.

It is well know from many mobility studies, such as the one developed by Google~\cite{google-mobility}, that in the countries where measures to reduce the mobility have been implemented, the effective reductions range between $20\%$ and $60\%$. Therefore, we must observe the interval $\alpha\in[0.4,0.8]$. Clearly, in the case above 10 beds per thousand people the situation is under control with a minimal reduction of mobility. However, for the case below 5 beds per thousand people, $\alpha>0.6$ results in SHFR greater than $0.5$, implying that the severe patients have the same mortality than critical patients. Hence, the Infected Fatality Rate reaches the value of $7.5\%$ approximately.

\section{Discussion}

The mobility parameter $\alpha(t)$ is a way to introduce a reduction of the number of interactions from the baseline random mixing giving by $S\frac{I}{N}$. The mobility should be extended to incorporate the population density distribution, strong but specific deviations from average mobility, and the way people from different areas move to other areas. 

In future work we will incorporate all these mobility aspects, and we consider important to incorporate the role that several other variables have in the system under study. For example, information technologies will have in the control of the disease by not only using AI to predict and detect trends, but also to incorporate tele-medicine to the possible treatments~\cite{yang2020modified,yang2020modified}. Moreover, it is important to consider how other non-health related variables are impacted by the pandemics such as economical drops~\cite{toda2020susceptible} and social demands. 
\section{Conclusion}
We conclude that mobility must remain below to $0.4$ to control the pandemics without a hospitalization system, and when there is a hospitalization system the mobility can increase above $0.4$, but only up to a limit which is specified by the number of beds available, as shown in Fig.~\ref{SHFRStatic}. Using this model it is possible to provide a much deeper understanding of how to control the pandemics by forecasting the pandemic situation including both the reductions of mobility of different partial or total lockdowns and the plan for increase of hospital capacity which most countries are applying.
\section{Acknowledgments}
The authors acknowledge the access to supercomputing capacity from the National Laboratory for High Performance Computing (NLHPC), Universidad de Chile, Powered@NLHPC. Partial economic support is acknowledged to Programa de Apoyo a Centros con Financiamiento Basal AFB 170004 to FCV, Instituto Milenio Centro Interdisciplinario de Neurociencia de Valparaíso (ICM-ANID P09-022-F). Research was partially sponsored by the Army Research Laboratory under Cooperative Agreement Number W911NF-19-2-0242, and by the Air Force Office of Scientific Research under award number FA9550-19-1-0368. The views and conclusions contained in this document are those of the authors and should not be interpreted as representing the official policies, either expressed or implied, of the Army Research Laboratory or the U.S. Government. The U.S. Government is authorized to reproduce and distribute reprints for Government purposes notwithstanding any copyright notation herein. Any opinions, finding, and conclusions or recommendations expressed in this material are those of the author(s) and do not necessarily reflect the views of the United States Air Force.
\bibliographystyle{ieeetr}
\bibliography{biblio}

\end{document}